\documentstyle[12pt]{article}

\def\sspace{\baselineskip = .16in}

\def\dspace{\baselineskip = .30in}

\def\VEV#1{\left\langle #1\right\rangle}

\newcounter{eqsave}
\def\subeqs{\refstepcounter{equation}
\setcounter{eqsave}{\value{equation}}
\setcounter{equation}{0}
\def\theequation{\arabic{eqsave}\alph{equation}}}
\def\endsubeqs{\setcounter{equation}{\value{eqsave}}
\def\theequation{\arabic{equation}}}

\begin{document}

\begin{titlepage}
\begin{flushright} BA-94-46\\
UMD-PP-94-99\\
hep-ph/9409381 \\
\end{flushright}
\vspace{0.9cm}
\begin{center} {\LARGE A Hint From the Inter-Family \\\vspace{10pt}
Mass Hierarchy: Two Vector-Like Families\\\vspace{10pt}
in the TeV Range}\\\vspace{40pt}
{\large K.S. Babu\footnote{Work supported in part by a grant from the DOE}}
\vspace{0.4cm}

{\it Bartol Research Institute\\ University of Delaware\\ Newark, DE 19716}
\vspace{0.5cm}

{\large Jogesh C. Pati\footnote{Work supported in part by a grant from the
NSF}}
\vspace{0.4cm}

{\it Department of Physics\\ University of Maryland\\ College Park, MD 20742}
\vspace{0.5cm}

{\large Hanns Stremnitzer}
\vspace{0.4cm}

{\it Institute for Theoretical Physics\\
University of Vienna\\
Boltzmanngasse 5, A-1090\\ Vienna, Austria}
\end{center}
\end{titlepage}

\sspace
\newpage
\begin{abstract}
Two vector--like families $Q_{L,R} = (U,D,N,E)_{L,R}$ and
$Q^\prime_{L,R} =$\newline $(U^\prime,D^\prime,N^\prime,E^\prime)_{L,R}$ with
masses of order 1 TeV, one of which is a doublet of $SU(2)_L$ and the
other a doublet of $SU(2)_R$, have been predicted to exist, together with
the three observed chiral families, in the context of a viable and
economical SUSY preon model.  The model itself possesses many attractive
features which include explanations of the origins of (i) diverse
mass--scales, (ii) family--replication, (iii) protection of the masses
of the composite quarks and leptons compared to their compositeness
scale and (iv) inter--family mass--hierarchy.  The existence of the two
vector--like families -- a prediction of the model -- turns out to be
crucial especially for an explanation of the inter--family
mass--hierarchy (IFMH).  Given the simplicity of the
explanation, the observed IFMH in turn appears to us to be a strong hint in
favor of the existence of the two vector--like families.

This paper is devoted to a detailed study of the expected masses,
mixings and decay modes of the fermions belonging to the two
vector--like families, in the context of the SUSY preon model, with the
inclusion of the renormalization--effects due to the standard model
gauge interactions.  Including QCD renormalization--effects, the masses
of the vector--like quarks are expected to lie in the range of 500 GeV to
about 2.5 TeV, while those of the vector--like leptons are expected to
be in the range of 200 GeV to 1 TeV.  Their mass pattern and decay modes
exhibit certain distinguishing features and
characteristic signals.  For example, when the LHC and, possibly
a future version of the SSC are built, pair--production of the
vector--like quarks would lead to systems such as ($b\overline{b} + 4Z+W^+W^-$)
and ($b\overline{b}+2Z+W^+W^-$), while an $e^-e^+$ linear collider (NLC)
of suitable energy can produce appreciably a single neutral heavy lepton $N$
together with $\nu_\tau$, followed by the decay of $N$ into
$(Z+\nu_\tau) \rightarrow (e^-e^+) + \nu_\tau$.  This last signal
may conceivably materialize even at LEP 200 if $N$ is lighter than about 190
GeV
\end{abstract}
\newpage
\dspace

\section*{I. Introduction}

Searches for the Higgs bosons and supersymmetry -- related directly and/or
indirectly to electroweak symmetry breaking -- are among the commonly
cited motivations, deservedly so, for the building of the hadronic
colliders like the LHC, a future version of the now extinct SSC,
as well as the next $e^-e^+$ linear collider NLC.
The discovery of the Higgs boson(s) will clearly shed
light on the problem of electroweak symmetry breaking and thereby on the
origin of the masses of $W$ and $Z$, while that of supersymmetry will
provide assurance on the common understanding of the gauge--hierarchy
problem.  But by themselves, none of these discoveries
would shed
light on an understanding of the inter--family mass--hierarchy (IFMH) and
therefore on the origin of the masses of the quarks and the leptons, of
which all matter is made.  As
regards this last issue, while there are a few explanations, we believe
that there is a particularly simple one which deserves attention.  For
this simple explanation of the IFMH to hold, there must exist
two ``vector--like'' quark--lepton families
$Q_{L,R}$  and $Q^\prime_{L,R}$ with masses of order 1 TeV,
where $Q_{L,R}$ couple vectorially
--i.e., in a parity conserving manner--to $W_L$'s, while $Q^\prime_{L,R}$
couple vectorially (assuming a left--right symmetric gauge theory) to
$W_R$'s.

As it turns out, two such vector--like families and the
associated fermion mass matrix, providing an explanation of the IFMH, arise in
a compelling manner in a supersymmetric preon model [1-3].
Since a search for these two vector--like families need
facilities like the SSC and the LHC, by way of emphasizing the dire need
for the building of such accelerators, we
first recall the {\it essential role} which
these two families play in providing an explanation of the inter--family
mass hierarchy.  The main purpose of the paper is to spell out the
expected properties of these two families -- i.e., their masses, mixings
and decay modes -- in some detail.  These should facilitate their search,
if and when the LHC, a possible new version of the SSC
and/or the NLC are built.

Before proceeding further, it is useful to recall one crucial
distinction between the chiral and the two vector--like families.
Since $Q_L$ and $Q_R$ couple symmetrically to $SU(2)_L
\times U(1)_Y$ gauge bosons, the mass term $(\overline{Q}_LQ_R+h.c)$ and
likewise $(\overline{Q}_L^{\prime}Q_R^{\prime}+h.c)$ preserve $SU(2)_L
\times U(1)_Y$.  In a class of models, the masses of these vector--like
families, protected by additional symmetries (see below),
turn out naturally to be
of order 1 TeV, rather than being of order Planck mass or some other
superheavy scale.  Because their masses are $SU(2) \times
U(1)$--symmetric, however, the oblique parameters $S,T$ and $U$ [4], or
equivalently $\rho$ (or $\epsilon_1$) and $\epsilon_3$ [5] do not
receive contributions from these vector--like families, in the leading
approximation.  As a result, the prevailing set of measurements of the
electroweak parameters, despite their precision, are not sensitive enough
to the existence of vector--like families [6].
This is unlike the case of a fourth chiral family which
is slightly disfavored and a technicolor family which seems to be excluded (at
least in its simple form) by the measurement of the $S,T$ and
$U$--parameters.  This leads one to infer that
if new families beyond the three are yet to be
found, they are more likely to exhibit vectorial rather than chiral
couplings to $W_L$'s and $W_R$'s (at least in their canonical forms
before $Q-Q^{\prime}$ mixings) [6].

The reason why we take the possible existence
of two vector--like families $Q_{L,R} = (U,D,N,E)_{L,R}$ and
$Q^{\prime}_{L,R}=(U^\prime,D^\prime,N^\prime, E^\prime)_{L,R}$ with
masses of order 1 TeV seriously is two--fold.
First of all, as mentioned above, they arise as a {\it
compelling prediction} of a SUSY composite model [1],
which seems to possess many attractive features.
These include an understanding of (a) the origin of family replication
[2], (b) protection of the masses of composite quarks and leptons,
compared to their compositeness scale [7], and (c) the origin of
the diverse mass scales--from
$M_{\rm Planck}$ to $m_\nu$ [1].  In addition, the model provides several
testable predictions [1-3], by which it can be excluded, if it is
wrong.

Second of all, {\it the existence of the two vector--like families is
found to be crucial in the model
to the very origin of the masses of the three chiral
families and simultaneously, to an understanding of the
inter--family mass--hierarchy}.  Both features turn out to
have their origin [1,3] in a spontaneously induced see--saw pattern for
the $5 \times 5$ mass matrix of the three chiral and the two
vector--like families, in which the direct mass terms of the three
chiral families $m^{(0)}(q_L^i \rightarrow q_R^j)$ vanish naturally
owing to underlying symmetries of the theory, barring
small corrections that are less than or
of order 1 MeV.  Although the compositeness scale is
determined on various grounds to be around $10^{11}~GeV$, owing to
protection by a chiral symmetry (which is the non--anomalous
$R$--symmetry),
the two vector--like families turn
out to acquire masses of order only 1 TeV.
The chiral families acquire masses primarily only by their
mixings with the two vector--like families.  One {\it general
consequence} of such a mass--matrix, which follows simply from its rank,
(see details later),
is that {\it one linear combination of the three chiral
families, is guaranteed to remain massless, barring corrections of order
1 MeV}, which is thus identified with the electron family.  At the same
time, the heaviest chiral fermion (top) acquires a mass of nearly
100-170 GeV and the masses of the fermions belonging to the muon family lie
intermediate in the range of 100-1500 MeV [3].  In this way, the see-saw
mass pattern of the type generated in the SUSY composite model provides
s simple resolution of the puzzle of inter--family mass--hierarchy.  In
particular, it explains the large hierarchy between $m_e$ and $m_t$.
Since such a pattern would not be possible without the two vector--like
families, the observed inter--family hierarchy seems to be
a {\it strong hint} in favor of the
existence of two such families with masses of order 1 TeV.  (In the
sequel, it will be clear as to why their number will have to be
precisely two--no less and no more).

Due to
the mixing of the chiral with the two vector--like
families the model suggests [3,1] a sizeable strength of
$\Delta m(D-\overline{D})$ and observable rates for $\mu \rightarrow
3e$ and especially for $t \rightarrow Zc$.  Such mixings also lead to a
lengthening of the tau lifetime and simultaneously to a correlated
small decrease in the
LEP neutrino counting $N_\nu$ from 3 [6].  Barring these small indirect
effects, these new families can, of
course, be discovered only provided machines like the SSC and the LHC
and possibly TeV--range $e^+e^-$ colliders are built.

With this in view, we spell out certain
characteristics of the two vector--like families--i.e., their expected
masses, mixings and decay modes in the context of
Ref. 1-3.
These turn out to possess some rather unexpected and interesting
features because of the see--saw pattern of the fermion mass--matrix on
the one hand and the differing renormalization group effects on the
entries in the mass matrix involving $Q_{L,R}$ as opposed to those
involving $Q^{\prime}_{L,R}$, on the other hand.  The differences in the
renormalization effects arise because $Q_{L,R}$ couple to $W_L$ and
$Q^\prime_{L,R}$ do not.  Among the characteristic signals, we find that
the production and subsequent decays of certain pairs of the heavy
quarks would give rise to systems such as $b \overline{b} + 4Z + 2W$ and
$b \overline{b} + 2Z + 2W$.  With the $Z$--boson decaying via charged
lepton pairs, these decay modes would provide distinctive signals and
would thereby facilitate the search for these new vector--like families,
at future hadronic colliders.

An additional intriguing signal is the single production of a neutral
heavy lepton ($N$) together with $\nu_\tau$ in an $e^+e^-$ machine of
appropriate energy, followed by the decay of of $N$ into $(\nu_\tau+Z)$
$\rightarrow (\nu_\tau + e^+e^-$).  This last signal may even be visible
at LEP 200 if $N$ is lighter than about 190 GeV.

\section*{II. Masses and Mixings of the Vector--Like Families}

The masses of the five--family system -- three chiral $(q_{L,R}^i$) and
two vector--like ($Q_{L,R}$ and $Q^{\prime}_{L,R}$) -- arise as follows.
(For details we refer the reader to Ref. 1 and 3.)

It is assumed that as the asymptotically free metacolor force becomes
strong at a scale $\Lambda_M \sim 10^{11}~GeV$ (this corresponds to an
input effective coupling $\overline{\alpha}_M \simeq 1/27$ near the
Planck scale), it forms a few SUSY--preserving and also SUSY--breaking
condensates.  Among the latter are the fermionic condensates consisting
of the metagaugino pair $\VEV{\vec{\lambda}.\vec{\lambda}}$ and the
preonic fermion pairs $\VEV{\overline{\psi}^f \psi^f}_{f=u,d}$ and
$\VEV{\overline{\psi}^c\psi^c}_{c=(r,y,b),l}$, where $f$ and $c$ denote
the flavor and the color attributes of the preons [1].  Noting that
in the model under consideration, a dynamical breaking of SUSY would be
forbidden in the absence of gravity, owing to the Witten index
theorem [8],
it has been argued that each of these fermion condensates which happen to
break SUSY, must be damped  by $(\Lambda_M/M_{Pl})$ [7], so that they would
vanish in the absence of gravity (i.e., in the limit of $M_{Pl}
\rightarrow \infty$).  Thus, one has:
\begin{eqnarray}
\VEV{\overline{\psi}^a\psi^a} & = & a_{\psi_a}\Lambda_M^3(\Lambda_M/M_{Pl})
\nonumber \\
\VEV{\vec{\lambda}.\vec{\lambda}} & = & a_{\lambda}\Lambda_M^3(\Lambda_M/
M_{Pl})
\end{eqnarray}
$a=u,d,r,y,b,l$.  The coefficients $a_{\lambda}$ and $a_{\psi_a}$ are
apriori expected to be of order unity, although $a_{\lambda}$ is
expected to be bigger than $a_{\psi_a}$ by a factor of 3 to 10 (say),
because $\lambda$'s are in the adjoint and $\psi$'s in the fundamental
representation of the metacolor group.  One can argue that even a
bosonic condensate $<\phi^* \phi>$, if it forms, would be damped by
powers of $(\Lambda_M/M_{Pl})$ [7].

Given these fermionic condensates, the vector--like families $Q$ and
$Q^{\prime}$ acquire relatively heavy Dirac masses
$O[a_{\lambda}\Lambda_M(\Lambda_M/M_{Pl})] \sim O(1~TeV)$ through the
metagaugino condensate $\VEV{\vec{\lambda}.\vec{\lambda}}$, which does
not distinguish between up and down flavors and between quarks and leptons.
Owing to extra symmetry associated with the composite chiral families
($q_{L,R}^i$), however, their direct mass--terms $m^{(0)}_{\rm dir}(q_L^i
\rightarrow q_R^j)$ cannot be induced through either $\VEV{\vec{\lambda}
.\vec{\lambda}}$ or $\VEV{\overline{\psi}\psi}$.  These receive small
contributions through e.g., products of chiral symmetry breaking
$\VEV{\overline{\psi}\psi}$ and bosonic $\VEV{\phi^*\phi}$ condensates,
which are thus damped by $(\Lambda_M/M_{Pl})^2$ and are $ \le O(1~MeV)$.

The chiral families (especially the $\mu$ and the $\tau$--families),
acquire their
masses almost entirely through their off--diagonal
$mixings$ with the vector--like
families, which are induced by the $\VEV{\overline{\psi}\psi}$--
condensates.  As a result, the mass matrices for the five--family system
(barring electroweak corrections and $m^{(0)}_{dir} \le 1~ MeV$) have the
following form at $\Lambda_M$ [1-3]:
\begin{eqnarray}
M_{f,c}^{(o)} =
\bordermatrix{& q_{L}^{i} & Q_{L} & Q_{L}^{\prime}\cr
\overline{q_{R}^{i}} & O & X\kappa_{f} & Y\kappa_{c}\cr
\overline{Q_{R}} & Y^{\prime\dagger}\kappa_{c} & \kappa_{\lambda} & O\cr
\overline{Q_{R}^{\prime}} & X^{\prime\dagger}\kappa_{f} & O &
\kappa_{\lambda}\cr}~~~.
\end{eqnarray}
Here $f=u$ or $d$ and $c = (r,y,b)$ or $l$, thus this form of the mass
matrix applies to four sectors $q_u, q_d, l$
and $\nu$.  The superscript $i=1,2,3$ runs over three chiral families.
The entries $X, Y, X^{\prime}$ and $Y^{\prime}$ are column
matrices in the space of these three chiral families
with entries of order unity.  In the
above $\kappa_{\lambda} \equiv O(a_{\lambda})\Lambda_M(\Lambda_M/M_{Pl})
\sim O(1 ~ TeV)$; $\kappa_{f,c} \equiv O(a_{\psi_{f,c}}) \Lambda_M (\Lambda_M/
M_{Pl}) = [O(a_{\psi_{f,c}})/O(a_{\lambda})]\kappa_{\lambda}$.  Following
remarks made above, we expect
\newline $\kappa_{f,c} = O(1/3~to~1/10)
\kappa_{\lambda}$.  {\it Thus the Dirac mass matrices of all four sectors
have a natural see--saw structure}, and the two vector--like families
acquire masses $\simeq \kappa_{\lambda} \sim O(1~TeV)$, while the three
chiral families acquire lighter masses $\sim (X_iY_i^{\prime *}+
X_i^{\prime *}Y_i)(\kappa_f \kappa_c/\kappa_{\lambda}) \ll
\kappa_{\lambda}$ at $\Lambda_M$ [3].

In the absence of electroweak corrections, which are typically of order
(1--10)\%, we have $X=X^{\prime}$ and $Y=Y^{\prime}$.  Furthermore,
the same $X,Y$ and $\kappa_{\lambda}$ apply to all four sectors: $q_u,
q_d, l$ and $\nu$.  A still further reduction in effective parameters is
achieved by rotating the chiral fields $q_L^i$ and $q_R^i$ so that
without loss of generality the row matrices $Y^T = Y^{\prime T}$ can be
brought to the simple form $(0, 0, 1)$ and $X^T = X^{\prime T}$ can be
brought to the form
$(0, p, 1)$ with redefined $\kappa_f$ and $\kappa_c$.  One can argue
plausibly on the basis of preon--diagram that $p$ is less than one, but
not very much smaller than one--i.e, $p \approx 1/2~to~1/10$ is
reasonable.

Given these rotated forms of $X$ and $Y$, {\it it is clear that one family is
massless, barring corrections of order 1 MeV} which arise from the direct
mass terms.  This one is identified with the electron--family.  The
masses of the muon--family, evaluated at $\Lambda_M$ are given by
$(m_{``\mu{\rm ''}})^{f,c} \simeq (p^2/2)(\kappa_f \kappa_c/\kappa_\lambda)$,
while those of the tau family are given by $(m_{``\tau{\rm ''}})^{f,c} \simeq
2(\kappa_f \kappa_c/\kappa_\lambda)$.  The $\mu/\tau$ mass ratio is thus
given by $p^2/4$.  For a value of $p \approx 1/3~{\rm to}~1/4$ (say),
which is not too small and reasonable, one thus obtains a large
hierarchy in the $\mu/\tau$ mass-ratio of about 1/40 to 1/64, as
observed.  In this way, the $5 \times 5$ see-saw mass-matrix, with the
approximate zeros dictated by the symmetry of the theory, provides a
very simple explanation of
the inter--family hierarchy: $m_{u,d,e} \ll m_{c,s,\mu} \ll
m_{t,b,\tau}$ with $m_e \sim O(1 MeV)$ and $m_t \sim O(100 GeV)$.  In
particular, it explains why $(m_e/m_t)$ is small ($\sim 10^{-5}$).
The role of the two vector--like families is, of course, crucial to
obtain such an explanation.  {\it This is the reason why we expressed in the
introduction that the two vector--like families may hold the key to an
understanding of the inter--family hierarchy.}

In the present note, we are primarily concerned with the masses, mixings
and decay modes of the two heavy families.  For this it is necessary to
retain their mixing at least with the tau family.  For the sake of
simplicity, we will ignore their mixings with the lighter electron and the muon
families.  This is, of course, a very good approximation, especially for the
electron family.  Even for the muon family, the $\mu-Q$ mixing angle is
smaller than the $\tau-Q$ mixing angle by the factor of $p \approx 1/3~
{\rm to}~1/4$ [3].

Thus, ignoring the mixings with the
$e$ and the $\mu$--families, which are defined only after the
transformation of the $X$ and $Y$--matrices to the simple forms as
mentioned above, the truncated $3 \times 3$ mass matrix of the tau and
the two vector--like families, evaluated at $\Lambda_M$, is given by

\begin{eqnarray}
\hat{M}_{f,c}^{(o)} =
\bordermatrix{& q_{L}^{\tau} & Q_{L} & Q_{L}^{\prime}\cr
\overline{q_{R}^{\tau}} & O & \kappa_{f} & \kappa_{c}\cr
\overline{Q_{R}} & \kappa_{c} & \kappa_{\lambda} & O\cr
\overline{Q_{R}^{\prime}} & \kappa_{f} & O &
\kappa_{\lambda}\cr}~~.
\end{eqnarray}

\noindent Noting that we expect $(\kappa_{f,c}/\kappa_\lambda) =
O(1/5~ {\rm to}~1/10)$, if we block--diagonalize this matrix to remove
the $q-Q$ and $q-Q^\prime$ entries, the $\tau$--family would acquire a
mass $\approx 2 \kappa_f \kappa_c/\kappa_\lambda$ (as mentioned before),
while the $2 \times 2$ sector involving $(Q,Q^\prime)$ would have a
symmetrical mass-matrix with two equal diagonal elements
$(\kappa_\lambda+ (\kappa_f^2+\kappa_c^2)/2\kappa_\lambda)$ and
off-diagonal elements $(\kappa_f \kappa_c/\kappa_\lambda)$ (to leading
order in $\kappa_{f,c}/\kappa_\lambda$).  From this, it would appear
that the two heavy eigenstates would be given essentially by
$(Q\pm Q^\prime)/\sqrt{2}$ corresponding to maximal mixing (barring
small admixtures of tau family in each case).  This is, however, greatly
distorted due to electroweak renormalizations which distinguish between
$Q_{L,R}$ and $Q^\prime_{L,R}$, to which we now turn.

We evaluate the running of the electroweak and QCD
coupling constants $g_{1,2,3}$ in the regime spanning
from $\Lambda_M \approx 3 \times
10^{11}~GeV$ to 1.5 TeV, using
renormalization--group equations.  In this regime,
3 chiral and two vector--like
families, the two Higgs--like multiplets and their superpartners
contribute to the $\beta$--functions.  We use
the familiar expressions for the mass--corrections--i.e. [9],
\begin{equation}
[m(Q)/m(\Lambda_M)] = \Pi [g_i^2(Q)/g_i^2(\Lambda_M)]^{b_i^m/2b_i}
\end{equation}
where $b_i$ and $b_i^m$ are the coefficients that appear in the
$\beta_i$ and $\gamma_i$ functions respectively,
\begin{eqnarray}
b_i & = & {1 \over 16 \pi^2} \left[-{11 \over 3} t_2(V) + {2 \over 3} t_2(F)
+ {1 \over 3} t_2(S)\right]  \\
b_i^m & = & {3 \over 8 \pi^2} C_2(R).k
\end{eqnarray}
with $C_2(R) = (N^2-1)/(2N)$ for $SU(N)$, while $C_2(R)=y^2$ for $U(1)$,
where $y$ is the normalized hypercharge.  The factor $k$ is 2/3 (1) for
a SUSY (non-SUSY) gauge theory.
The $b_i$'s are given by
\begin{equation}
b_3 = {1 \over {16 \pi^2}},~ b_2 = {5 \over {16 \pi^2}},~b_1 = {(53/5) \over
{16 \pi^2}}
\end{equation}
\begin{equation}
b_3^m = {1 \over {4 \pi^2}}\left({4 \over 3}\right),~ b_2^m = {1 \over {
4 \pi^2}} \left({3 \over 4}\right) ~~~.
\end{equation}
$b_1^m$ can take seven different values depending upon the hypercharges
of the external pairs of fermions.  Denoting these seven values of
$b_1^m$ by $\hat{a}_j
\equiv (a_j/4 \pi^2)$~$(j=1-7)$, the values of the $a_j$'s for the
relevant pairs of fermions are listed below:
\begin{eqnarray}
a_1(\overline{U}_Rt_L, \overline{U}_RU_L, \overline{D}_Rb_L,
\overline{D}_RD_L)
 & = & {1 \over 6}.{1 \over 6}.{3 \over 5} = {1 \over 60} \nonumber \\
a_2(\overline{t}_RU_L^\prime,\overline{U}^\prime_RU_L^\prime) & = & {2 \over
3}.{2 \over 3}. {3 \over 5} = {4 \over 15} \nonumber \\
a_3(\overline{U}_R^\prime t_L, \overline{b}_RD_L^\prime,
\overline{D}_R^\prime D_L^\prime) & = &
\left({-1 \over 3}\right),\left({-1 \over 3}\right).{3 \over 5} =
{1 \over 15} \nonumber \\
a_4(\overline{b}_RD_L,\overline{D}_R^\prime b_L) & = &
{1 \over 6}.\left({-1 \over 3}\right) {3 \over 5} = -{1 \over 30}
\nonumber \\
a_5(\overline{E}_R\tau_L,\overline{N}_R\nu_{\tau L},
\overline{N}_RN_L, \overline{E}_RE_L) & = & \left({-1 \over 2}\right).
\left({-1 \over 2}\right).{3 \over 5} =
{3 \over 20} \nonumber \\
a_6(\overline{\tau}_RE_L,\overline{E}_R^\prime\tau_L) & =&
\left({-1 \over 2}\right).\left({-1}\right).{3 \over 5} = {3 \over 10}
\nonumber \\
a_7(\overline{E}^\prime_RE^\prime_L) & = &
\left({-1}\right).\left({-1}\right).{3 \over 5} = {3 \over 5}
\end{eqnarray}
The factor $3/5$  is the usual normalization
factor for the hypercharge $y$.

The mass renormalization group
parameters $\eta_c, \eta_L, \eta_{1,2,3,4,5,6,7}$, (corresponding to
$SU(3)^C,~SU(2)_L$ and $U(1)_Y$) are defined
as
\begin{eqnarray}
\eta_c &=& \left[{{g_3^2(1.5~TeV)} \over
{g_3^2(\Lambda_M)}}\right]^{-b_3^m/2b_3} \nonumber \\
\eta_L &=& \left[{{g_2^2(1.5~TeV)}\over {g_2^2(\Lambda_M)}}\right]^{-b_2^m
/2b_2} \nonumber \\
\eta_i &=& \left[{{g_1^2(1.5~TeV)} \over {g_1^2(\Lambda_M)}}\right]^{-
\hat{a}_i/
2b_1},~ i=1-7~.
\end{eqnarray}
To evaluate these factors, we choose as
input values
$g_1^2(M_Z) = 0.2136,~g_2^2(M_Z) = 0.4211,~g_3^2(M_Z)=1.4828$,
corresponding to the values of
$\alpha=1/127.9, {\rm sin}^2\theta_W = 0.2333$ and
$\alpha_3 = 0.118$ measured at $M_Z$ at LEP [10].
With three chiral and two vector--like families along with two Higgs
multiplets and all their superpartners
contributing to the $\beta$ and $\gamma$ functions in the energy
regime 1.5 TeV to $\Lambda_M$, the gauge couplings at
$\Lambda_M = 3 \times 10^{11}~GeV$ are
evaluated to be $g_1^2(\Lambda_M) = 0.5163,~g_2^2(\Lambda_M) = 0.8192,
{}~g_3^2(\Lambda_M) = 1.5958$ [11].  Using (7)-(10), the
resulting $\eta$ factors are thus given by
\begin{eqnarray}
\eta_c &=& 2.39,~\eta_L = 1.23,~\eta_1=1.003,~\eta_2=1.043 \nonumber \\
\eta_3 &=& 1.011,~\eta_4=0.995,~\eta_5=1.024,~\eta_6=1.049,~\eta_7=1.1.
\end{eqnarray}
As expected, QCD renormalization effect denoted by $\eta_c$ is large.
That due to $SU(2)_L$, denoted by $\eta_L$, is significant.
Note, however, that the renormalization factors due to the
hypercharge interaction given by $\eta_1-\eta_7$ are typically rather
small, differing from unity by 1 to 10\%.

In addition to the renormalization group running from $\Lambda_M$ to the
weak scale, the mass parameters will also receive QCD and electroweak
radiative corrections at $\Lambda_M$, which lead to differences between
$X$ and $X^\prime$ and between $Y$ and $Y^\prime$ (see eq. (2)).
These corrections are typically
of the order of 5-10\%, which are
to be compared with the renormalization effects due to running given in
eq. (11).  Note that the $SU(2)_L$ running factor $\eta_L$ is much
bigger than these preonic corrections at $\Lambda_M$, while the
hypercharge running corrections are of similar magnitude.  In what
follows, we shall neglect the corrections arising from the preon
diagrams, which should represent a good approximation for the masses and
mixings involving the top family and the vector--like families.  (These
corrections at $\Lambda_M$ do play an important role for the masses and
mixings involving the $e$ and the $\mu$ families, see Ref. 3.)

Inserting the renormalization factors into (3),
the $3 \times 3$ sector of the up, down and charged lepton mass matrices
involving the tau and the two vector--like families take the form
\begin{eqnarray}
M_{up} = \bordermatrix{& t_L & U_L & U_L^\prime \cr
\overline{t}_R & 0 & \kappa_u \eta_3 & \kappa_r
\eta_2 \cr
\overline{U}_R & \kappa_r \eta_L
\eta_1 & \kappa_{\lambda} \eta_L \eta_1
& 0 \cr
\overline{U}_R^\prime &
\kappa_u \eta_3 & 0 & \kappa_{\lambda}  \eta_2\cr} \eta_c
\end{eqnarray}

\begin{eqnarray}
M_{down} = \bordermatrix{& b_L & D_L & D_L^\prime \cr
\overline{b}_R & 0 & \kappa_d  \eta_4 & \kappa_r
\eta_3 \cr
\overline{D}_R & \kappa_r \eta_L \eta_1 & \kappa_{\lambda} \eta_L
\eta_1 & 0 \cr
\overline{D}_R^\prime &
\kappa_d  \eta_4 & 0 & \kappa_{\lambda}  \eta_3 \cr} \eta_c
\end{eqnarray}

\begin{eqnarray}
M_{lepton}^+ = \bordermatrix{& \tau_L & E_L & E^\prime_L \cr
\overline{\tau}_R & 0 & \kappa_d \eta_6 & \kappa_l \eta_7 \cr
\overline{E}_R &
\kappa_l \eta_L \eta_5 & \kappa_{\lambda} \eta_L \eta_5 & 0 \cr
\overline{E}_R^\prime & \kappa_d
\eta_6 & 0 & \kappa_{\lambda} \eta_7\cr}~.
\end{eqnarray}

In the neutral lepton sector, since $\nu_{\tau_R}$ becomes
superheavy ($m_{\nu_{\tau R}} \sim \Lambda_M \sim 10^{11}~GeV$) [1,3,12], the
corresponding mass matrix is a $2 \times 3$ matrix given by
\begin{eqnarray}
M_{lepton}^0 = \bordermatrix{& \nu_{\tau L} & N_L & N_L^\prime \cr
\overline{N}_R & \kappa_l \eta_L \eta_5 & \kappa_\lambda
\eta_L \eta_5 & 0 \cr
\overline{N}_R^\prime & \kappa_u & 0 & \kappa_\lambda\cr}~~.
\end{eqnarray}

Since $\kappa_{f,c} \ll \kappa_{\lambda}$,
the mass matrices of eq. (12-15) can be diagonalized by using the see--saw
formula.  The light mass eigenvalues corresponding to the chiral quarks
and leptons of the tau family are given by
\begin{eqnarray}
m_t & \simeq & 2{{\kappa_u \kappa_r} \over {\kappa_{\lambda}}} \eta_c \eta_3
,~~~~ m_b \simeq 2{{\kappa_d \kappa_r}\over {\kappa_{\lambda}}}
\eta_c \eta_4 \nonumber \\
m_\tau & \simeq & 2{{\kappa_d \kappa_l}\over {\kappa_{\lambda}}} \eta_6
,~~~~m_{\nu_\tau} \simeq 0~~.
\end{eqnarray}

The $2 \times 2$ mass matrix corresponding to the heavy vector--like
fermions in the up sector
is given by
\begin{eqnarray}
M_H^U =\bordermatrix{& U_L & U^\prime_L \cr
\overline{U}_R & \eta_L \eta_1(1+{{\kappa_r^2}\over{2\kappa_\lambda^2}}) +
{{\kappa_u^2}\over{2 \kappa_\lambda^2}}{{\eta_3^2}\over{\eta_1\eta_L}}
& {{\kappa_u \kappa_r} \over {2 \kappa_\lambda^2}} \eta_3 \left(
{{\eta_L \eta_1} \over {\eta_2}}+{{\eta_2}\over {\eta_1 \eta_L}}\right) \cr
\overline{U}^\prime_R &
{{\kappa_u \kappa_r}\over {\kappa_\lambda^2}} \eta_3
& \eta_2(1+{{\kappa_r^2}\over{2 \kappa_\lambda^2}})+{{\kappa_u^2}\over
{2 \kappa_\lambda^2}}{{\eta_3^2}\over{\eta_2}}}\eta_c \kappa_\lambda
\end{eqnarray}

Although the basis vectors in (17) are slightly different from those in
(12), we use the same symbols.
The analog mass--matrix for the down sector of the heavy fermions
is obtained from the above by the
replacement $(\eta_3 \rightarrow \eta_4,~\eta_2 \rightarrow \eta_3,~
\kappa_u \rightarrow \kappa_d)$.  The charged lepton matrix is obtained
from eq. (17) by the replacement $(\eta_c \rightarrow 1,~\kappa_r
\rightarrow \kappa_l, \kappa_u \rightarrow \kappa_d, \eta_1 \rightarrow
\eta_5, \eta_2 \rightarrow \eta_7, \eta_3 \rightarrow \eta_6)$.  In the
neutral lepton sector, the squared mass matrix corresponding to the
$(N_R, N_R^{\prime})$ fields is given by
\begin{eqnarray}
M_{lepton}^0 M_{lepton}^{0^\dagger} = \left(\matrix{(\kappa_l^2+
\kappa_\lambda^2)\eta_L^2\eta_5^2 & \kappa_l\kappa_u\eta_L\eta_5 \cr
\kappa_l\kappa_u\eta_L\eta_5 & \kappa_\lambda^2+\kappa_u^2}\right)
\end{eqnarray}

The mass eigenvalues of the heavy fermions are obtained from the above:
\begin{eqnarray}
M_{U_1} & \simeq & \kappa_\lambda\eta_c\left[\eta_L\eta_1(1+{{\kappa_r^2}
\over{2\kappa_\lambda^2}})+{{\eta_3^2}\over {\eta_1\eta_L}}{{\kappa_u^2}
\over {2 \kappa_\lambda^2}}\right] \nonumber \\
M_{D_1} & \simeq & \kappa_\lambda\eta_c\left[\eta_L\eta_1(1+{{\kappa_r^2}
\over{2\kappa_\lambda^2}})+{{\eta_4^2}\over {\eta_1\eta_L}}{{\kappa_d^2}
\over {2 \kappa_\lambda^2}}\right] \nonumber \\
M_{U_2} & \simeq & \kappa_\lambda\eta_c\left[\eta_2(1+{{\kappa_r^2}
\over{2\kappa_\lambda^2}})+{{\eta_3^2}\over {\eta_2}}{{\kappa_u^2}
\over {2 \kappa_\lambda^2}}\right] \nonumber \\
M_{D_2} & \simeq & \kappa_\lambda\eta_c\left[\eta_3(1+{{\kappa_r^2}
\over{2\kappa_\lambda^2}})+{{\eta_4^2}\over {\eta_3}}{{\kappa_d^2}
\over {2 \kappa_\lambda^2}}\right] \nonumber \\
M_{N_1} & \simeq & \kappa_\lambda \eta_L \eta_5 \left[1+{{\kappa_u^2}\over
{2 \kappa_\lambda^2}}{{\eta_L^2\eta_5^2}\over{\eta_L^2\eta_5^2-1}}
-{{\kappa_l^2}\over{2\kappa_\lambda^2}}{1 \over{\eta_L^2\eta_5^2-1}}
\right] \nonumber \\
M_{E_1} & \simeq & \kappa_\lambda\left[\eta_L\eta_5(1+{{\kappa_l^2}
\over{2\kappa_\lambda^2}})+{{\eta_6^2}\over {\eta_5\eta_L}}{{\kappa_d^2}
\over {2 \kappa_\lambda^2}}\right] \nonumber \\
M_{N_2}  & \simeq &  \kappa_\lambda [1-{{\kappa_u^2}\over
{2 \kappa_\lambda^2}}{{1}\over{\eta_L^2\eta_5^2-1}}
+{{\kappa_l^2}\over{2\kappa_\lambda^2}}{{\eta_L^2\eta_5^2}
\over{\eta_L^2\eta_5^2-1}}] \nonumber \\
M_{E_2} & \simeq & \kappa_\lambda\left[\eta_7(1+{{\kappa_l^2}
\over{2\kappa_\lambda^2}})+{{\eta_6^2}\over {\eta_7}}{{\kappa_d^2}
\over {2 \kappa_\lambda^2}}\right]~~.
\end{eqnarray}

It is easy to verify that $U_1$ and $D_1$ contain primarily the
$SU(2)_L$--doublet $Q$--fermions with a small admixture of
$SU(2)_R$--doublets $Q^\prime$, while $U_2$ and $D_2$ contain primarily
$Q^\prime$ with a small admixture of $Q$.
Since $M_H^{U,D,E}$ are not symmetric, there will be in
general two mixing angles, one for the left--handed sector and one for
the right--handed sector.  Note, however, that if one neglects the small
corrections due to hypercharge interactions $\eta_1 - \eta_7$ and
keeps only the lowest order
terms in $\epsilon_L$,  where
$\eta_L \equiv 1+\epsilon_L$, the above matrices become symmetrical.  In this
approximation (which should be close to the true scenario, since the
neglected terms are only of order 5\% or so), the (symmetric) mixing
angles for the charged fermion sectors are given by
\begin{eqnarray}
{\rm tan}2\theta_{U,D} & \simeq &
{{2\kappa_{u,d}\kappa_r}\over {\kappa_\lambda^2
(\eta_L-1)}} \nonumber \\
{\rm tan}2\theta_E & \simeq &
{{2 \kappa_d\kappa_l}\over{\kappa_\lambda^2(\eta_L-1
)}}~~.
\end{eqnarray}
The mass eigen--states $(U_1)_{L,R}$ and $(U_2)_{L,R}$ are given by
\begin{eqnarray}
(U_1)_{L,R} &=& {\rm cos}\theta_U U_{L,R} + {\rm sin}\theta_U
U^\prime_{L,R} \nonumber \\
(U_2)_{L,R} &=& -{\rm sin}\theta_U U_{L,R} + {\rm cos}\theta_U U^\prime_{L,R}
\end{eqnarray}
and likewise $(D_{1,2})_{L,R}$ and $(E_{1,2})_{L,R}$.

Note that the mixing in the up--sector of the heavy quarks denoted by
$\theta_U$
is proportional to $\kappa_u$ and
hence is large, while in the down sector, the mixing angle $\theta_D$
and $\theta_E$ are proportional to
$\kappa_d$; they are consequently small (since $\kappa_d/\kappa_u \approx
m_b/m_t \ll 1$).

The mixing angle for the right--handed neutral leptons $(N_R, N^\prime_R
$) is
\begin{equation}
{\rm tan}2\theta_N^R \simeq {{2 \kappa_u\kappa_l\eta_L}\over{\kappa_\lambda^2
(\eta_L^2-1)}}~~.
\end{equation}

In the left--handed neutral lepton sector, the mixing is somewhat more
complicated, due to the fact that $\nu_{\tau_L}-N_L$ mixing is not
negligible.  Starting from eq. (15), we obtain the exact orthogonal
matrix $V_L^\nu$ which transforms the gauge eigenstate ($\nu_L^g$)
to the mass eigenstate
via $\nu_L^m = V_L^\nu \nu_L^g$, where
\begin{eqnarray}
V_L^\nu = \left(\matrix{c_1 & -c_2s_1 & -s_1s_2 \cr c_3s_1 & c_1c_2c_3+s_2s_3
& c_1c_3s_2-c_2s_3 \cr -s_1s_3 & -c_1c_2s_3+c_3s_2 & -c_2c_3-c_1s_2s_3}
\right)~~.
\end{eqnarray}
Here $s_i = {\rm sin}\theta_i, c_i={\rm cos}\theta_i$ with
\begin{eqnarray}
{\rm tan}\theta_1 & =& {{\sqrt{\kappa_u^2+\kappa_l^2}}\over
{\kappa_\lambda}}~,~~~~ {\rm tan}\theta_2 = {{\kappa_u}\over{\kappa_l}}
\nonumber \\
{\rm tan}2\theta_3 & = & {{2\kappa_u\kappa_l\kappa_\lambda
\sqrt{\kappa_\lambda^2+\kappa_u^2+\kappa_l^2}(\eta_L^2\eta_5^2-1)}\over
{(\kappa_u^2+\kappa_l^2\eta_L^2\eta_5^2)(\kappa_l^2+\kappa_u^2+
\kappa_\lambda^2)-\kappa_\lambda^2(\kappa_l^2+\eta_L^2\eta_5^2\kappa_u^2
)}} \nonumber \\
& \simeq & {{2 \kappa_u\kappa_l}\over{\kappa_l^2-\kappa_u^2}}~~.
\end{eqnarray}
Since $\kappa_{u,l} \ll \kappa_\lambda$, the mixing angle
$\theta_1$ is small ($O(1/5~{\rm to}~1/10$), say);
but $\theta_2$ and $\theta_3$ can be relatively
larger, since $\kappa_u$ and $\kappa_l$ are expected to be comparable
within a factor of two (say).  Note that $\theta_2 \simeq \theta_3$.

To have a feel for the numerical values of the masses and the mixing
angles that are relevant to the heavy families, we need to know the
effective parameters which enter into the fermion mass--matrix.  Talking
of these, note that the full $5 \times 5$ mass--matrices (eq. (2)) of
the four sectors -- i.e., $u,d,l$ and $\nu$ -- which in general could
involve 100 parameters, even if they are all real, are essentially
determined (barring electroweak corrections and $m^{(0)}_{\rm dir}$
which are important only for the light families) by just six effective
parameters: $p, \kappa_u, \kappa_d, \kappa_r, \kappa_l$ and
$\kappa_\lambda$.  These successfully describe the gross features of the
masses of the known fermions -- in particular their magnitudes and the
inter--family hierarchy [3].  In fact, even these few parameters are not
completely arbitrary (unlike in the case of an elementary
Higgs--picture) in that we know apriori their approximate values to
within a factor of 10, say.  For example, as mentioned above [1,3],
we expect that $\kappa_\lambda = {\cal O}(a_\lambda \Lambda_M
(\Lambda_M/M_{Pl})) \sim {\cal O}(1~TeV)$ and that $(\kappa_{f,c}/
\kappa_\lambda) \approx 1/3$ to $1/10$, while $1/10 < p < 1/2$ (say).

Guided in part by the expected order of magnitude of these parameters
and their ratios as mentioned above and by the observed masses of the
muon and the tau--families, we are led to the following values for
certain ratios of these parameters:
\subeqs
\begin{eqnarray}
(\kappa_r/\kappa_l) & \approx & 0.9~{\rm to}~1 \\
(\kappa_d/\kappa_u) & \approx & {1 \over {40 \pm 10}} \\
(\kappa_l/\kappa_\lambda) & \approx & 1/3 \\
{{\kappa_u \kappa_r}\over {\kappa_\lambda}} & \approx & (26 ~{\rm to}
{}~33)~GeV~~~.
\end{eqnarray}
These values are arrived at as follows [See Ref. 3, a more detailed
discussion will be presented elsewhere].  The
ratio $(\kappa_r/\kappa_l)$ is obtained from $(m_b/m_\tau)$ including
QCD and electroweak renormalization effects, and $(\kappa_d/\kappa_u)$ is
obtained from $(m_b/m_t)$ (see below).  The ratio
$(\kappa_l/\kappa_\lambda)$ is obtained utilizing corrections of order
$(\kappa_l/\kappa_\lambda)^2$ to the leading see--saw contribution to
$m_\tau$ (see Eq. (7) of Ref. 3) and optimizing the choice of parameters
consistent with their expected ranges so as to obtain a reasonable fit
to $(m_\mu/m_\tau)$ within 10\%, with the inclusion of electroweak
corrections [Ref. 3].  Finally, the ratio $(\kappa_u \kappa_r/
\kappa_\lambda)$ given by (25-d) is obtained by using the fact that
internal consistency of the model requires $m_t(phys) \stackrel{_<}
{_\sim} 160~GeV$ [3,13].

Using (25-a) and (25-c) we obtain
\begin{equation}
(\kappa_r/\kappa_\lambda) \approx 0.33
\end{equation}
which, combined with (25-d), yields $\kappa_u \approx (76-100)~GeV$.
Allowing for some uncertainty in $(\kappa_l/\kappa_\lambda)$ which
affects $\kappa_u$, we take
\begin{equation}
\kappa_u \approx (40-100)~GeV~~~.
\end{equation}
This still leaves $\kappa_\lambda$ undetermined (although its order of
magnitude is known).  We observe that the ``number'' of
light neutrinos measured at LEP places an upper
limit on $(\kappa_u/\kappa_\lambda)$.  Owing to $\nu_\tau-N^\prime$
mixing, the model yields [6]: $N_\nu = 2+ [1-2(\kappa_u/
\kappa_\lambda)^2]$.  Comparing with the observed value [10] $N_\nu = 2.99
\pm 0.03$, we get $(\kappa_u/\kappa_\lambda) \stackrel{_<}{_\sim} 1/7$
for $N_\nu \stackrel{_>}{_\sim} 2.96$.  Allowing for agreement within
two standard deviations on the one hand and following our
general expectations for
$(\kappa_u/\kappa_\lambda)$ mentioned above, on the other hand, we take:
\begin{equation}
(\kappa_u/\kappa_\lambda) \approx 1/5 ~{\rm to}~ 1/10~~.
\end{equation}
Values of $(\kappa_u/\kappa_\lambda)$ near $1/10$ can be probed if
$N_\nu$ can be measured with an accuracy of 0.01 to 0.02.  For the
present, using (25-f and g), we get
\begin{eqnarray}
\kappa_\lambda  & \approx & (1~{\rm to}~2)(200-500)~GeV \nonumber \\
& \approx & (200~GeV-1~TeV)~.
\end{eqnarray}
\endsubeqs

To have a feel for the masses and mixing angles, consider a
representative set of values:
\begin{eqnarray}
\kappa_\lambda & \approx & 575~GeV,~\kappa_u/\kappa_\lambda = 1/6,~
\kappa_d/\kappa_u \simeq 1/40 \nonumber \\
& ~ & \kappa_r/\kappa_\lambda \approx 1/3,~
\kappa_l/\kappa_\lambda \approx 1/3
\end{eqnarray}
This yields [9,13]
\begin{eqnarray}
(m_t, m_b, m_\tau)_{1.5~TeV} \simeq (135, 3.4, 1.5)~GeV \nonumber \\
(m_t, m_b, m_\tau)_{phys} \simeq (157, 4.7, 1.7)~GeV
\end{eqnarray}
\begin{eqnarray}
U_1 \simeq 1814~ GeV ~~~~~N_1 \simeq 765~GeV \nonumber \\
D_1 \simeq 1787 ~GeV ~~~~~E_1 \simeq 763~GeV \nonumber \\
U_2 \simeq 1504 ~GeV ~~~~~N_2 \simeq 581~GeV \nonumber \\
D_2 \simeq 1465 ~GeV ~~~~~E_2 \simeq 667~GeV
\end{eqnarray}
\begin{eqnarray}
\theta_U & \simeq & 1/5,~\theta_D \simeq 1/210,~\theta_E \simeq 1/140
\nonumber \\
\theta_N^R & \simeq & 1/10.6,~{\rm tan}\theta_1 \simeq 1/2.7,~{\rm tan}
\theta_2 \simeq 1/2~,{\rm tan}\theta_3 \simeq 1/2.6~~~.
\end{eqnarray}
There are corrections of order 10\% to these numbers arising from the
mixing of the top and vector--like families with the lighter $e$ and
$\mu$ families, but they can be neglected for our present purpose.

While the precise values of the masses and the mixing angles depend upon
the specific choice of parameters given by (26), which in general could
vary within the range indicated in (25 a-h), a few qualitative features
would still remain which are worth noting:

(1)  It is clear that the mixing angles $\theta_D$ and $\theta_E$ are
tiny ($\sim 10^{-2})$ while those in the up--sector, including
neutrino--members, are appreciable.  This is simply because
$\kappa_d/\kappa_u \sim 1/40$ and thus $\kappa_d/\kappa_\lambda
\stackrel{_<}{_\sim}1/240$.

(2)  Given that $\theta_D \ll 1$ and $\theta_U \approx 1/5$,
it follows from (21)
that $D_1$ and even $U_1$ are mostly composed of $Q$--fermions which are
$SU(2)_L$--doublets while $U_2$ and $D_2$ are mostly composed of
$Q^\prime$--fermions which are $SU(2)_R$--doublets.  This is important
for their decay modes.

(3)  Note (from (19) and (28))
that the pair $U_1$ and $D_1$ are nearly degenerate to within
about 10-30 GeV, so also the pair $N_1$ and $E_1$, and to a lesser
extent the pair
$N_2$ and $E_2$.  But the $(U_1,D_1)$ pair is substantially heavier, by
about a few hundred GeV, than the pair $(U_2,D_2)$.  Similarly the
$(N_1,E_1)$ pair is heavier by about 100 GeV than the pair $(N_2,E_2)$.
This is because the $(U_1,D_1)$ and also the $(N_1,E_1)$ pair receive
enhancement due to $SU(2)_L$--renormalization factor $\eta_L$, which is,
however, absent for the $(U_2,D_2)$ and $(N_2,E_2)$--pairs.

The mass--gap between the quark--pairs $(U_1,D_1)$ versus $(U_2,D_2)$ is
much larger than that between the leptonic pairs $(N_1,E_1)$ versus
$(N_2,E_2)$ because the QCD--enhancement factor $\eta_c \approx 2.39$
multiplies $\eta_L \approx 1.23$ in the case of quarks, but not for the
leptons.

All these qualitative features of the heavy fermion mass spectrum remain
intact even if the relevant parameters are varied within the range
indicated in eq. (25).

(4)  Given this mass--pattern, we see that $U_1 \rightarrow D_1 + W$ and
likewise $N_1 \rightarrow E_1 + W$ are forbidden kinematically, while
decays such as $U_1 \rightarrow D_2 +W,~U_1 \rightarrow U_2+Z,~
D_1 \rightarrow U_2+W,~D_1 \rightarrow D_2+Z$ and possibly
$U_2 \rightarrow D_2+W$ are kinematically allowed.

\section*{III. Coupling and Decay Modes of the Heavy Fermions}

In terms of the mixing angles in the heavy sector,
the coupling of the $W_L^{\pm}$ and
$Z^0$ to the left--handed as well as the right--handed charged fermions
can be written down.  We list them in Tables 1-5 presented below.

\noindent {\bf Table. 1.}  The coupling of $W_L^{\pm}$ to the
left--chiral quarks of the $\tau$ and the two vector--like families.
An overall factor $(g/\sqrt{2})
\gamma_\mu$ is not displayed, but should be understood.
\vskip .1in
\begin{center}
\begin{tabular}{|c||c|c|c|}\hline
{}~ & $\overline{t}_L$ & $\overline{U}_{1L}$ & $\overline{U}_{2L}$ \\
\hline
$b_L$ & $1-(\kappa_u^2+\kappa_d^2)/ (2\kappa_{\lambda}^2)$
& $-(\kappa_u/\kappa_{\lambda}) {\rm sin}\theta_U$ &
$(\kappa_u/ \kappa_{\lambda}) {\rm cos}\theta_U$ \\ \hline
$D_{1L}$ & $-(\kappa_d/ \kappa_{\lambda}){\rm sin}\theta_D$ &
cos$\theta_U$cos$\theta_D$ & sin$\theta_U$cos$\theta_D$ \\ \hline
$D_{2L}$ & $(\kappa_d/ \kappa_{\lambda}) {\rm cos}\theta_D$ &
cos$\theta_U$sin$\theta_D$ & sin$\theta_U$sin$\theta_D$ \\
\hline
\end{tabular}
\end{center}
\vskip0.2in

\noindent {\bf Table. 2.}  The coupling of $W_L^{\pm}$ to the
right--chiral quarks of the tau and the vector--like families.
An overall factor
$(g/\sqrt{2})
\gamma_\mu$ is not displayed.
\vskip .1in
\begin{center}
\begin{tabular}{|c||c|c|c|}\hline
{}~ & $\overline{t}_R$ & $\overline{U}_{1R}$ & $\overline{U}_{2R}$ \\
\hline
$b_R$ & $\kappa_u \kappa_d/(\kappa_\lambda^2 \eta_L^2)$
& $-(\kappa_d/\kappa_{\lambda}\eta_L) {\rm cos}\theta_U$ &
$-(\kappa_d/ \kappa_{\lambda}\eta_L) {\rm sin}\theta_U$ \\ \hline
$D_{1R}$ & $-(\kappa_u/\kappa_{\lambda}\eta_L){\rm cos}\theta_D$ &
cos$\theta_U$cos$\theta_D$ & sin$\theta_U$cos$\theta_D$ \\ \hline
$D_{2R}$ & $-(\kappa_r/ \kappa_{\lambda}) {\rm sin}\theta_D$ &
cos$\theta_U$sin$\theta_D$ & sin$\theta_U$sin$\theta_D$ \\
\hline
\end{tabular}
\end{center}
\vskip0.2in

\noindent {\bf Table. 3.}  The coupling of $Z^0$ to the
left--chiral quarks and charged leptons collectively denoted by
$q_f$ and $Q_f$, respectively, where $i=1,2$, $f=$up, down, charged
lepton.  An overall factor $-(g/
{\rm cos}\theta_W) \gamma_\mu$ is not displayed.  Here $T_3= (1/2, -1/2,
-1/2)$ and $Q=(2/3, -1/3, -1)$ for $f=u,d,l$, $s_W^2={\rm sin}^2
\theta_W$ and $F=U,D,E,N$.
\vskip .1in
\begin{center}
\begin{tabular}{|c||c|c|c|}\hline
{}~ & $\overline{q}_L^f$ & $\overline{Q}_{1L}^f$ & $\overline{Q}_{2L}^f$ \\
\hline
$q_L^f$ & $T_3(1-\kappa_f^2/\kappa_\lambda^2)-Qs_W^2$ & $-T_3(\kappa_f/
\kappa_\lambda) {\rm sin}\theta_F$ & $T_3 (\kappa_f/\kappa_\lambda)
{\rm cos}\theta_F$ \\ \hline
$Q_{1L}^f$ & $-T_3 (\kappa_f/\kappa_\lambda){\rm sin}\theta_F$ &
$T_3{\rm cos}^2\theta_F -Q s_W^2$ & $T_3 {\rm cos}\theta_F {\rm sin}
\theta_F$  \\ \hline
$Q_{2L}^f$ & $T_3(\kappa_f/\kappa_\lambda){\rm cos}\theta_F$ &
$T_3 {\rm cos}\theta_F {\rm sin}\theta_F$ & $T_3 {\rm sin}^2\theta_F
-Qs_W^2$ \\
\hline
\end{tabular}
\end{center}
\vskip0.2in

\noindent {\bf Table. 4.}  The coupling of $Z^0$ to the
right--chiral quarks and charged leptons with the same notation as in
Table 3.
\vskip .1in
\begin{center}
\begin{tabular}{|c||c|c|c|}\hline
{}~ & $\overline{q}_R^f$ & $\overline{Q}_{1R}^f$ & $\overline{Q}_{2R}^f$ \\
\hline
$q_R^f$ & $T_3 (\kappa_f^2/\kappa_\lambda^2\eta_L^2)-Qs_W^2$ & $-T_3(\kappa_f/
\kappa_\lambda \eta_L) {\rm cos}\theta_F$ & $-T_3 (\kappa_f/\kappa_\lambda)
{\rm sin}\theta_F$ \\ \hline
$Q_{1R}^f$ & $-T_3 (\kappa_f/\kappa_\lambda\eta_L){\rm cos}\theta_F$ &
$T_3{\rm cos}^2\theta_F -Q s_W^2$ & $T_3 {\rm cos}\theta_F {\rm sin}
\theta_F$  \\ \hline
$Q_{2R}^f$ & $-T_3(\kappa_f/\kappa_\lambda){\rm sin}\theta_F$ &
$T_3 {\rm cos}\theta_F {\rm sin}\theta_F$ & $T_3 {\rm sin}^2\theta_F
-Qs_W^2$ \\
\hline
\end{tabular}
\end{center}
\vskip0.2in

\noindent {\bf Table. 5.}  The coupling of $Z^0$ to the
left--chiral neutral leptons.   Notation same as in Table 3. The overall
factor $(-g/{\rm cos}\theta_W)\gamma_\mu$ is not displayed.
\vskip .1in
\begin{center}
\begin{tabular}{|c||c|c|c|}\hline
{}~ & $\overline{\nu}_{\tau L}$ & $\overline{N}_{1L}$ & $\overline{N}_{2L}$ \\
\hline
$\nu_{\tau L}$ & $1-s_1^2s_2^2$ & $s_1s_2(c_1c_3s_2-c_2s_3)$ & $-s_1s_2(
c_2c_3+c_1s_2s_3)$ \\
\hline
$N_{1L}$ & $s_1s_2(c_1c_3s_2-c_2s_3)$ &
$1-(c_1c_3s_2-c_2s_3)^2$ & $(c_1c_3s_2-c_2
s_3)(c_2c_3+c_1s_2s_3)$
\\ \hline $N_{2L}$ &
$-s_1s_2(c_2c_3+c_1s_2s_3)$ & $(c_1c_3s_2-c_2s_3)(c_2c_3+c_1s_2s_3)$ &
$1-(c_2c_3+c_1s_2s_3)^2$
\\
\hline
\end{tabular}
\end{center}
\vskip0.2in

The coupling of $W_L^+$ to the
left--handed charged lepton currents
take the following form:
\begin{eqnarray}
\overline{\nu_\tau}\tau_L & : & 1-{{\kappa_u^2+\kappa_d^2}\over{2
\kappa_\lambda^2}} \nonumber \\
\overline{\nu_\tau}E_{1L} & : & -{{\kappa_d}\over{\kappa_\lambda}}{\rm sin}
\theta_E \nonumber \\
\overline{\nu_\tau}E_{2L} & : & {{\kappa_d}\over{\kappa_\lambda}}
{\rm cos}\theta_E \nonumber \\
\overline{N}_{1L}\tau_L & : & c_3 s_1 -(c_2c_3+s_2s_3){{\kappa_l}\over
{\kappa_\lambda}} \nonumber \\
\overline{N}_{1L}E_{1L} & : & c_3s_1\left({{\kappa_l}\over{\kappa_\lambda}}
{\rm cos}\theta_E-{{\kappa_d}\over{\kappa_\lambda}}{\rm sin}\theta_E
\right) +
(c_1c_2c_3+s_2s_3)\left({\rm cos}\theta_E(1-{{\kappa_l^2}\over
{2 \kappa_\lambda^2}})+{\rm sin}\theta_E{{\kappa_l\kappa_d}\over
{2 \kappa_\lambda^2}}\right) \nonumber \\
\overline{N}_{1L}E_{2L} & : & c_3s_1({{\kappa_l}\over{\kappa_\lambda}}
{\rm sin}\theta_E+{{\kappa_d}\over{\kappa_\lambda}}{\rm cos}\theta_E)+
(c_1c_2c_3+s_2s_3)\left({\rm sin}\theta_E(1-{{\kappa_l^2}\over
{2 \kappa_\lambda^2}})-{\rm cos}\theta_E{{\kappa_l\kappa_d}\over
{2 \kappa_\lambda^2}}\right) \nonumber \\
\overline{N}_{2L}\tau_L & : & -s_1s_3-(-c_1c_2s_3+c_3s_2){{\kappa_l}\over
{\kappa_\lambda}} \nonumber \\
\overline{N}_{2L}E_{1L} & : & -s_1s_3({{\kappa_l}\over{\kappa_\lambda}}
{\rm cos}\theta_E-{{\kappa_d}\over{\kappa_\lambda}}{\rm sin}\theta_E)+
(-c_1c_2s_3+c_3s_2)\left({\rm cos}\theta_E(1-{{\kappa_l^2}\over
{2 \kappa_\lambda^2}})+{\rm sin}\theta_E{{\kappa_l\kappa_d}\over
{2 \kappa_\lambda^2}}\right) \nonumber \\
\overline{N}_{2L}E_{2L} & : & -s_1s_3({{\kappa_l}\over{\kappa_\lambda}}
{\rm sin}\theta_E+{{\kappa_d}\over{\kappa_\lambda}}{\rm cos}\theta_E)+
(-c_1c_2s_3+c_3s_2)\left({\rm sin}\theta_E(1-{{\kappa_l^2}\over
{2 \kappa_\lambda^2}})-{\rm cos}\theta_E{{\kappa_l\kappa_d}\over
{2 \kappa_\lambda^2}}\right) \nonumber \\
\end{eqnarray}

Using Tables 1-5 and the mixing angles listed, the following pattern of
decay modes  for the heavy fermions emerge (Table 6).
We list all dominant modes
and only some suppressed or forbidden ones.  The reason for suppression
or dominance of a mode can be inferred by looking at the third column.
Rates which are proportional to sin$^2\theta_D$ and/or
$(\kappa_d/\kappa_\lambda)^2$ are suppressed.

\newpage
\noindent {\bf Table. 6.} Pattern of heavy fermion decay modes.  We have
approximated the leptonic mixing angles $\theta_{1,2,3}$  of Eq. (24)
by keeping only the lowest order terms in $\kappa_l/\kappa_\lambda$ and
$\kappa_u/\kappa_l$.
\vskip .1in
\begin{center}
\begin{tabular}{|c|c|c|c|}\hline
Particle & Decay Modes & Rate ($\propto$) & Comment \\
{}~ & ~ & ~ & ~ \\ \hline\hline
$U_1$ &  $\rightarrow D_1+W^+$ & Kin. Forbidden & ~ \\
{}~     &  $\rightarrow D_2+W^+$ & $($cos$\theta_U$sin$\theta_D)^2$ &
Suppressed
\\
{}~ & $\rightarrow U_2+Z$ & (cos$\theta_U$sin$\theta_U)^2$ & Dominant \\
{}~ & $\rightarrow
t_L+Z$ & $((\kappa_u/2\kappa_\lambda){\rm sin}\theta_U)^2$ & Appreciable
\\
{}~ & $\rightarrow t_R+Z$ & $((\kappa_u/2\kappa_\lambda \eta_L)
{\rm cos}\theta_U)^2$ & Appreciable \\
\hline
$U_2$ & $\rightarrow D_1+W^+$ & Kin. Forbidden & ~ \\
{}~ & $\rightarrow b_L+W^+$ &
$((\kappa_u/\kappa_\lambda){\rm cos}\theta_U)^2$ & Dominant \\
{}~ & $\rightarrow t_L+Z$ & $(( \kappa_u/2\kappa_\lambda){\rm cos}
\theta_U)^2$ & Appreciable \\
{}~ & $\rightarrow t_R + Z$ & $((\kappa_u/2\kappa_\lambda){\rm sin}\theta_U)^2$
&
Appreciable \\
\hline
$D_1$ & $\rightarrow U_{2}+W^-$ & (sin$\theta_U {\rm cos}\theta_D)^2$ &
Dominant-1 \\
{}~ & $\rightarrow t_L+W^-$ & $((\kappa_d/\kappa_\lambda){\rm sin}\theta_D)^2
$ & Highly suppressed \\
{}~ & $\rightarrow t_R+W^-$  & $((\kappa_u/\kappa_\lambda\eta_L)
{\rm cos}\theta_D)
^2$ & Dominant-2 \\
\hline
$D_2$ & $\rightarrow t_L +W^-$ & $((\kappa_d/\kappa_\lambda){\rm cos}
\theta_D)^2$ & Suppressed \\
{}~ & $\rightarrow t_R+W^-$ & $((\kappa_r/\kappa_\lambda){\rm sin}\theta_D)^2
$ & Dominant \\
{}~ & $\rightarrow b_L+Z$ & $((\kappa_d/2\kappa_\lambda){\rm cos}\theta_D)^2$
& Suppressed \\
{}~ & $\rightarrow b_R + Z$ & $((\kappa_d/2\kappa_\lambda){\rm sin}\theta_D)^2$
 & Suppressed \\
\hline
$N_1$ & $\rightarrow E_1+W$ & Kin. Forbidden & ~ \\
{}~ & $\rightarrow E_{2L}+W$ & $(\kappa_l\kappa_d/\kappa_\lambda^2
(\eta_L-1))^2$ & Highly suppressed \\
{}~ & $\rightarrow N_{2L}+Z$ & $(\kappa_u\kappa_l/(2\kappa_\lambda^2))^2$
& Phase space supp. \\
{}~ & $\rightarrow N_{2R}+Z$ & ($(1/2){\rm cos}\theta_U{\rm sin}\theta_U)^2$ &
Phase space supp. \\
{}~ & $\rightarrow \nu_\tau+Z$ & $((\kappa_u/\kappa_\lambda)
(\kappa_u^3/\kappa_l^3 - \kappa_u\kappa_l/2\kappa_\lambda^2))^2$ &
Dominant \\
{}~ & $\rightarrow \tau_L+W$ & $(\kappa_l/\kappa_\lambda)^2$ & Dominant
\\ \hline
$N_2$ & $\rightarrow E_1+W$ & Kin. Forbidden & ~ \\
{}~ & $ \rightarrow \tau_L+W$ & $(\kappa_u/\kappa_\lambda)^2$ & Dominant \\
{}~ & $\rightarrow \nu_\tau+Z$ & $(\kappa_u/\kappa_\lambda)^2$ & Dominant \\
\hline
$E_1$ & $\rightarrow \tau_R+Z$ & $((\kappa_d/2\kappa_\lambda\eta_L) {\rm cos}
\theta_E)^2$ & Supp. rate, but dom. mode \\
{}~ & $ \rightarrow \tau_L+Z$ & $((\kappa_d/2\kappa_\lambda){\rm sin}
\theta_E)^2$ & Doubly suppr. \\
\hline
$E_2$ & $\rightarrow \nu_\tau+W$ & $((\kappa_d/\kappa_\lambda){\rm cos}
\theta_E)^2$ & Suppr. rate but dom. mode \\
{}~ & $\rightarrow \tau_L+Z$ & $((\kappa_d/2\kappa_\lambda){\rm cos}
\theta_E)^2$ & Same \\
{}~ & $\rightarrow \tau_R+Z$ & $((\kappa_d/2\kappa_\lambda){\rm sin}
\theta_E)^2$ & Highly suppr. \\
\hline
\end{tabular}
\end{center}
\vskip.1in

In above, it is to be understood that for modes, where chirality is not
shown (e.g., $U_1 \rightarrow D_1+W^+$), either chirality has
the same amplitude.  It is interesting to note, however, that {\it there are
modes for which there is a strong preference for either the left or the
right chirality}.  For example, $D_1 \rightarrow t_L+W^-$ is highly
suppressed, but $D_1 \rightarrow t_R+W^-$ would have appreciable
branching ratio.  Likewise, $D_2 \rightarrow t_R+W^-$ is dominant, but
$D_2 \rightarrow t_L+W^-$ is suppressed.  Furthermore, $E_1
\rightarrow \tau_R+Z$ is dominant but $E_1 \rightarrow \tau_L+Z$ is
suppressed, while $E_2 \rightarrow \tau_L+Z$ is dominant and
$E_2 \rightarrow \tau_R+Z$ is suppressed.  Such interesting decay
patterns and the correlation with chirality are clearly features that
are intimately tied as much to the nature of the gauge couplings of the
canonical fields $Q_{L,R}$ and $Q^\prime_{L,R}$ as to the nature of the
fermion mass--matrix given by (2) and (3).  In this sense, they are
truly distinguishing features of the model.

\section*{IV. Production and Signals}

Production of the heavy quarks in pairs by hadronic colliders at SSC and LHC
energies has been studied in a number of papers [14].  These studies
typically yield production cross sections at $\sqrt{s}=
40~TeV$ as follows:

\noindent {\bf Table. 7.} Heavy quark pair production cross section at
$\sqrt{s} = 40 ~TeV$ for $U \overline{U}$.
The cross section $\sigma$ listed is in $nb$.
\vskip .1in
\begin{center}
\begin{tabular}{|c|c|c|c|c|c|c|}\hline
$m_U$ & 3 TeV & 2.4 TeV & 2 TeV & 1.5 TeV  & 1 TeV & 500 GeV \\ \hline
$\sigma(pp \rightarrow U\overline{U})$ & $2 \times 10^{-6}$ & $10^{-5}$
& $5 \times 10^{-5}$ & $3 \times 10^{-4}$ & $2.5 \times 10^{-3}$ &
$5 \times 10^{-2}$ \\ \hline
\end{tabular}
\end{center}
Assuming that a future version of the SSC will be built one day in the
near future, the production
cross section noted above would lead to about
$2.5 \times 10^4$ events per year for $m_U=1~TeV$, with a
luminosity of $10^{33} cm^{-2}s^{-1}$.

We now consider the likely signals of pair production of such heavy
quarks by considering their expected decay modes.

(1)
\begin{eqnarray}
pp,p\overline{p} & \rightarrow & U_1 + \overline{U}_1 \nonumber \\
U_1 & \rightarrow & U_2+Z;~ (U_2 \rightarrow \{b+W, {\rm or}, t+Z\};
t \rightarrow  b+W)
\nonumber \\
\overline{U}_1 & \rightarrow & \overline{U}_2+Z;~~ \overline{U}_2 \rightarrow
(\{\overline{b}+W, {\rm or}, \overline{t}+Z\};
\overline{t} \rightarrow \overline{b}+
W\})
\end{eqnarray}

Thus
\subeqs
\begin{eqnarray}
(pp, \overline{p}p) \rightarrow  U_1 + \overline{U}_1 & \rightarrow &
2Z+2W+b\overline{b} \\
{\rm or}~&\rightarrow & 4Z+2W+b\overline{b}\\
{\rm or}~& \rightarrow & 3Z+2W+b\overline{b}
\end{eqnarray}
\endsubeqs
(2)
\begin{eqnarray}
pp, \overline{p}p & \rightarrow & U_2+\overline{U}_2 \nonumber \\
U_2 & \rightarrow & \{b+W, {\rm or}, t+Z\}; t\rightarrow b+W \nonumber \\
\overline{U}_2 & \rightarrow & \{\overline{b}+W, {\rm or}, \overline{t}+Z\};
\overline{t} \rightarrow \overline{b}+W
\end{eqnarray}
Thus,
\subeqs
\begin{eqnarray}
(pp,\overline{p}p) \rightarrow U_2 + \overline{U}_2 & \rightarrow & 2W+b
\overline{b} \\
&{\rm or}~ \rightarrow & 2W+2Z+b\overline{b} \\
&{\rm or}~ \rightarrow & 2W+Z+b\overline{b}
\end{eqnarray}
\endsubeqs
(3)
\begin{eqnarray}
pp, \overline{p}p & \rightarrow & D_1 + \overline{D}_1 \nonumber \\
D_1 & \rightarrow & \{U_2+W^-, {\rm or}, t+W^-\}; t \rightarrow b+W \nonumber
\\
\overline{D}_1 & \rightarrow & \{\overline{U}_2+W, {\rm or}, \overline{t}+W\};
\overline{t} \rightarrow \overline{b}+W
\end{eqnarray}
Thus noting signals for $U_2\overline{U}_2$ in (33), we expect
\subeqs
\begin{eqnarray}
(pp,\overline{p}p) \rightarrow D_1 \overline{D}_1 & \rightarrow &
4W+b\overline{b} \\
& {\rm or} ~\rightarrow & 4W+2Z+b\overline{b}\\
& {\rm or}~\rightarrow & 4W+Z+b\overline{b}
\end{eqnarray}
\endsubeqs
(4)
\begin{eqnarray}
(pp, \overline{p}p) & \rightarrow & D_2 + \overline{D}_2 \nonumber \\
D_2 & \rightarrow & t+W, {\rm or}~ t \rightarrow b+W \nonumber \\
\overline{D}_2 & \rightarrow & \overline{t}+W, {\rm or},~
\overline{t} \rightarrow
\overline{b}+W \\
(pp, \overline{p}p) & \rightarrow & D_2\overline{D}_2 \rightarrow 4W+b
\overline{b}
\end{eqnarray}

Before discussing the signals for heavy lepton pair--production, we see
already from (31-38) that pair--production of the heavy quarks of the
model would lead to distinctive signatures -- such as production of
$4Z+W^+W^-+b\overline{b}$ --
 which would not be confused
first of all with the background expected from the standard model
involving the three families and the Higgs boson and second with the
signals expected from pair--production of some general heavy quark
belonging, for example, to a standard fourth family.

Heavy lepton pair production in hadronic and leptonic machines will
proceed through virtual photon and $Z^0$--productions.  These cross
sections have been studied in Ref [14].  The prominent decay modes of the
heavy leptons are listed in Table [6], from which we arrive at the
signals for heavy lepton production in the model.

(1)
\begin{eqnarray}
(p\overline{p},e^+e^-) \rightarrow N_1 + \overline{N}_1 \nonumber \\
N_1 \rightarrow \{\nu_\tau+Z, {\rm or},~ \tau+W\};
{}~~\tau \rightarrow e^-+\overline{\nu
} \nonumber \\
\overline{N}_1\rightarrow \{\overline{\nu}_\tau+Z, {\rm or},~
\overline{\tau}+W\}
;~~
\overline{\tau}\rightarrow e^++\nu
\end{eqnarray}
\subeqs
\begin{eqnarray}
p\overline{p} \rightarrow N_1\overline{N}_1 & \rightarrow & 2Z+\nu_\tau+
\overline{\nu}_\tau \\
&{\rm or}~ \rightarrow & 2 W+ \tau\overline{\tau}\\
&{\rm or}~ \rightarrow & Z+W + \overline{\tau}+\nu_\tau
\end{eqnarray}
\endsubeqs
(2)
\begin{eqnarray}
(p\overline{p},e^+e^-) & \rightarrow & N_2+\overline{N}_2 \nonumber \\
N_2 & \rightarrow & \{\tau+W, {\rm or},~ \nu_\tau+Z\} \\
(p\overline{p}, e^+e^-) & \rightarrow & \{2Z+\nu_\tau+\overline{\nu}_\tau;
2W+\tau+\overline{\tau}; W+Z+\overline{\tau}+\nu_\tau\}
\end{eqnarray}

(3)
\begin{eqnarray}
(pp,e^+e^-) & \rightarrow & E_1+\overline{E}_1 \nonumber \\
E_1 & \rightarrow & \tau_R+Z;~~ \overline{E}_1 \rightarrow \overline{\tau}_R+Z
\nonumber \\
(p\overline{p},e^+e^-) & \rightarrow & 2Z+\tau_R+\overline{\tau}_R
\end{eqnarray}

(4)
\begin{eqnarray}
(p\overline{p},e^+e^-) & \rightarrow & E_2+\overline{E}_2 \nonumber \\
E_2 & \rightarrow & \{\nu_\tau+W; \tau_L+Z\};~~ \overline{E}_2
\rightarrow \{\overline{\nu}_\tau+W; \overline{\tau}_L+Z\}
\nonumber \\
& \rightarrow & \{2W+\nu_\tau+\overline{\nu}_\tau; 2Z+\tau_L+\overline{\tau}_L;
W+Z+\nu_\tau +\overline{\tau}_L\}
\end{eqnarray}

Thus $N_1\overline{N}_1, N_2\overline{N}_2$ and $E_2\overline{E}_2$ lead
essentially to the same signals with three possible channels, while
$E_1\overline{E}_1$ leads only to $2Z+\tau_R+\overline{\tau}_R$.  These
may be compared with standard model signals arising from
$pp, p\overline{p},e^+e^- \rightarrow H+Z \rightarrow ZZZ
\rightarrow \tau^+\tau^-+2Z$.  One distinction which may not be easy to
utilize in practice is that the SM would yield $\tau^+\tau^-$ with
nearly equal left and right helicities for sin$^2\theta_W \simeq 1/4$
(since $J_\mu^Z \propto J^{3 em}_\mu-{\rm sin}^2\theta_W$), whereas the
model discussed here [1-3] yields predominantly
$\tau_R\overline{\tau}_R$.  The
other distinction has to do with expected rates and kinematics.

One additional feature mentioned in Chapter I is that one expects
significant production of a single heavy lepton ($N_1$ or $N_2$)
together with $\nu_\tau$ in an $e^+e^-$ machine of appropriate energy
through a virtual $Z$ (see Table 5 and eqs. (23), (24) and (29) for
relevant amplitudes and mixing angles) followed by the dominant decay of
$N_1$ (or $N_2$) into
$(\nu_\tau + Z) \rightarrow  \nu_\tau + e^+ e^-$;
\begin{eqnarray}
e^+ e^- & \rightarrow & ``Z{\rm ''} \rightarrow N_1 ~({\rm or}~ N_2) +
\nu_{\tau_L} \nonumber \\
& \rightarrow & (Z + \nu_{\tau_L}) + \nu_{\tau_L}
\rightarrow (e^+e^-)_{Z-mass} +
{}~{\rm missing~energy}
\end{eqnarray}
The amplitude for $``Z'' \rightarrow N_{2L} + \nu_{\tau L}$ (for example
) is $\sim s_1s_2 \sim (1/6 ~{\rm to}~1/10)$ relative to
$``Z'' \rightarrow e^+e^-$.  Thus the rates for such spectacular
signals, characterized by eq. (45) would be appreciable and observable.

\section*{V. Vector-Like Fermions in Other Models and Concluding Remarks}

Before concluding we wish to note that vector--like fermions have been
introduced in other contexts by several authors [15-18].  The case
considered here [1-3] possesses, however, some distinguishing features
as regards the origin and the nature of the vector--like families
compared to those arising in the context of other works [15-18].  Below we
present some of these distinctions.

In Refs. (15) and (16), $SU(2)_L$--singlet vector--like families, which
are in part analogous to our $Q^\prime_{L,R} = (U^\prime, D^\prime,
N^\prime, E^\prime)_{L,R}$, have been introduced to generate, with the
choice of suitable Higgs multiplets and/or discrete symmetries, a
see--saw like fermion mass--matrix.  A second motivation to introduce
vector--like families has been to account for the vanishing of the
strong CP $\overline{\theta}$--parameter at the tree-level [17].  A
related attempt introduces $SU(2)_L$--doublet vector--like quarks, but
not the corresponding leptons and the analog of $SU(2)_L$--singlet
$Q^\prime_{L,R}$, together with an extra $U(1)$--gauge symmetry to
arrange so that $\overline{\theta}$ is zero at the tree--level.  The
similarities and the differences between the vector--like fermions
arising in these models and those arising in the present case [1-3]
are listed below.

\noindent (i) First, the vector--like fermions or families are not
predicted in any of these models [15-18] by an apriori theoretical reason,
based on some higher symmetry or other grounds, whereas in the present
case, they arise in a compelling manner as a general consequence of
SUSY--compositeness in a class of SUSY theories based on QCD--like
binding force [1-3].

\noindent (ii) Second, the masses of the vector--like families, which
are $SU(2)_L \times U(1)$--invariant, are not protected by a symmetry in the
models of Ref. [15-18].
They could apriori be as large as the grand
unification or even Planck scale and thus inaccessible to future
accelerators.  By contrast, for the case considered here, the masses of
the vector--like families are protected by the non--anomalous
R symmetry, which is broken by the desired amount only by the
SUSY--breaking metagaugino condensate $<\lambda.\lambda>$.  Thus their
masses are protected to the same extent as supersymmetry breaking; both
are damped by the Planck scale and are naturally of order 1 TeV [1,7,3].
 This is what makes them accessible to accelerators.

\noindent (iii) Third, invariably the other models, as they stand,
contain only $SU(2)_L$--singlet family [16] or families [15], which are in
part analogous to our $SU(2)_R$--doublet family $Q^\prime_{L,R} =
(U^\prime, D^\prime, N^\prime, E^\prime)_{L,R}$, but none of them [15-18]
contain the analog of our complete $SU(2)_L$--doublet family
$Q_{L,R} = (U,D,N,E)_{L,R}$ together with the $SU(2)_L$--singlet
$Q^\prime_{L,R}$.  In short, the coupling of the vector--like fermions
to $W_L^\pm$ can serve to distinguish
between the cases of interest.

Vector--like fermions arise in a more compelling manner for the case of
a {\bf 27} of $E_6$, which can arise from the heterotic superstring
theory.  The {\bf 27} splits into ({\bf 16} + {\bf 10} + {\bf 1}) under
SO(10).  The {\bf 16} contains the standard model fermions and a
$(\nu_R)^c$.  The {\bf 1} gives a singlet neutral lepton $N_L$.  The
{\bf 10} contains a $(2_L, 2_R, 1^c)$ and a $(1_L, 1_R, 6^c)$ under
$SU(2)_L \times SU(2)_R \times SU(4)^c$ of $SO(10)$.  The $(1,1,6^c)$
gives a $SU(3)^c$-color triplet of quarks, with charge $-1/3$
($B-L = -2/3$), and an antitriplet with charge +1/3, which are {\it
singlets} of $SU(2)_L$, while the (2,2,1) gives a pair of leptonic
$SU(2)_L$--doublets, which couple vectorially to $SU(2)_L$ gauge
bosons.  These also form a pair of $SU(2)_R$--doublets.  These vector--
like leptons differ, of course, from the case considered here in that
they are not accompanied by vector--like quarks to make a complete
family.  Furthermore, they couple vectorially to $SU(2)_L$ as well as
$SU(2)_R$ gauge bosons.

In summary, two vector--like families, not more not less [19], with one
coupling vectorially to $W_L$'s and the other to $W_R$'s (before
mass--mixing), with masses of order 1 TeV, constitute a {\it hall--mark}
and a crucial prediction of the SUSY preon model [1-3].  {\it There does not
seem to be any other  model including superstring--inspired models of
elementary quarks and leptons
which have a good reason to predict two such
complete vector--like families with masses in the TeV range}.

To conclude, the simplicity with which the system of three chiral and
two vector--like families lead to the right gross pattern of the
inter--family mass--hierarchy and the fact that the presence of the two
vector--like families is fully compatible with the measurements of the
light neutrinos $N_\nu$ and of the oblique electroweak parameters
incline us to believe that the two
vector--like families may well exist in TeV region.  The observed
inter--family mass--hierarchy appears to be a strong hint in this
direction.  Establishing their absence in the TeV--region will clearly
exclude a class of preonic theories which are based on SUSY--QCD type
binding force.  On the other hand, their discovery, assuming especially
that their decay--pattern conforms with the one spelt out here, will
first of all
provide us assurance on the see--saw origin of the masses
of the observed quarks and leptons --i.e., on the validity of the
mass--matrix characterized by eq. (3).
At the same time, considering that no other
model seems to have a compelling reason to predict two vector--like
families with with masses in the TeV--range,
their discovery will clearly provide
a strong support to the preonic approach.  This in turn will provide
the much needed hint as to
how the superstring theories make contact with the low--energy world.

Thus we do hope that not only the LHC will be approved and built in the
near future but that efforts will continue and succeed to build a future
version of the SSC with $E_{cm}\ge 40~TeV$ and the NLC $e^+e^-$ machine
with $E_{cm} \approx 1~TeV$.  Without these, some very precious
discoveries, including the ones mentioned above, which are expected to
lie around the corner, will never materialize.

\section*{Acknowledgments}

The research of K.S.B is supported by a grant from the Department of
Energy, that of JCP
is supported in part by a grant from the National Science Foundation.
We wish to thank A.
Jawahery, J. Kim, T. Rizzo and G.A. Snow for many helpful discussions.  K.S.B
and H.S are thankful to the Elementary Particle
Theory Group of the University of Maryland
for the warm hospitality extended to them on several occasions
during the course of this work.

\section*{References}
\begin{enumerate}
\item J.C. Pati, Phys. Lett. {\bf B228}, 228 (1989).
\item K.S. Babu, J.C. Pati and H. Stremnitzer, Phys. Lett. {\bf B256},
206 (1991).
\item K.S. Babu, J.C. Pati and H. Stremnitzer, Phys. Rev. Lett.
{\bf 67}, 1688 (1991).
\item M. Peskin and T. Takeuchi, Phys. Rev. Lett. {\bf 65}, 964 (1990).
\item G. Altarelli, R. Barbieri and S. Jadach, Nucl. Phys. {\bf B 369},
3 (1992).
\item K.S. Babu, J.C. Pati and X. Zhang, Phys. Rev. {\bf D 46}, 2190
(1992).
\item J.C. Pati, M. Cvetic and H. Sharatchandra, Phys. Rev. Lett.
{\bf 58}, 851 (1987).
\item E. Witten, Nucl. Phys. {\bf B 185}, 513 (1981).
\item Considering that in this paper we are interested primarily in the
masses and the mixings of the vector--like families, we include here
only QCD and electroweak renormalization effects to evaluate the running
fermion masses, but neglect, for the sake of simplicity, radiative
effects due to off-diagonal and diagonal Yukawa couplings, which follow
the pattern of the fermion masses (see eqs. (2) and (3)).  The Yukawa
radiative effects can, of course, be important as regards predictions of
the masses of the lighter chiral families (eg. $m_t, m_b, m_\tau$ etc.),
especially if the Yukawa couplings are large (order unity).  But they
would not affect significantly the masses and mixings of the
vector--like families $relative$ to those of the chiral families,
because such relative effects can largely be simulated by choosing
$\kappa_\lambda$ appropriately within the range being considered (see
discussions later).
\item The LEP Electroweak Working Group, Preprint CERN-PPE/93-157
(August 93).
\item K.S. Babu and J.C. Pati, Phys. Rev. {\bf D 48}, R1921 (1993).
\item K.S. Babu, J.C. Pati and H. Stremnitzer, Phys. Lett. {\bf B264},
347 (1991).
\item The approximate upper limit of 150 GeV on $m_t(m_t)$, quoted in
Ref. 3, corresponds to an upper limit of about 157 GeV for $m_t(phys) =
m_t(m_t)[1+4\alpha_s/3\pi]$.  The question of whether the inclusion of
Yukawa radiative effects, which have been ignored in Ref. 3 [see remarks
in Ref. 9], can naturally allow the upper limit on $m_t(phys)$ to
increase by 10 to 20\% is being studied in collaboration with J. Kim.
This work will be reported in a forthcoming paper.
\item E. Eichten, I. Hinchliffe, K. Lane and C. Quigg, Rev. Mod. Phys.
{\bf 56}, 579 (1984)
\item Z. Berezhiani, Phys. Lett. {\bf 129B}, 99 (1983); D. Chang and
R.N. Mohapatra, Phys. Rev. Lett. {\bf 58}, 1600 (1987); A. Davidson and
K.C. Wali, Phys. Rev. Lett. {\bf 59}, 393 (1987); S. Rajpoot,
Phys. Lett. {\bf B191}, 122 (1987).
\item B.S. Balakrishna, Phys. Rev. Lett. {\bf 60}, 1602 (1988); B.S.
Balakrishna, A. Kagan and R.N. Mohapatra, Phys. Lett. {205B}, 345
(1988).
\item A.E. Nelson, Phys. Lett. {\bf 136B}, 387 (1984); S.M. Barr,
Phys. Rev. Lett. {\bf 53}, 329 (1984) and Phys. Rev. {\bf D30},
1805 (1984); K.S. Babu and R.N. Mohapatra, Phys. Rev. {\bf D41}, 1286
(1990).
\item P. Frampton and T. Kephart, Phys. Rev. Lett. {\bf 66}, 1666
(1991); E. Carlson and D. Land,HUTP-92/A008.
\item Having more than two vector--like families will, in general, spoil
one of the most desirable features of the fermion mass--matrix of the
five--family system which guarantees one zero eigenvalue [3] that
corresponds to the mass of the electron family.  Furthermore, subject to
left--right symmetry, if there are more than two vector--like families,
there would have to be four of them, i.e., two $Q_{L,R}$ and two
$Q^\prime_{L,R}$.  together with their SUSY partners, they would make
QCD coupling grow rapidly above 1 TeV to become confining below $10^{11}
$ GeV.  Thus, there appears to be a good reason why there should be
precisely two vector--like families, corresponding to $Q_{L,R}$ and
$Q^\prime_{L,R}$,no more no less.

\end{enumerate}
\end{document}